# A Generative Model Method for Unsupervised Multispectral Image Fusion in Remote Sensing

Arian Azarang, *Student Member, IEEE*, Nasser Kehtarnavaz, *Fellow, IEEE*

*Abstract*—**This paper presents a generative model method for multispectral image fusion in remote sensing which is trained without supervision. This method eases the supervision of learning and it also considers a multi-objective loss function to achieve image fusion. The loss function incorporates both spectral and spatial distortions. Two discriminators are designed to minimize the spectral and spatial distortions of the generative output. Extensive experimentations are conducted using three public domain datasets. The comparison results across four reduced-resolution and three full-resolution objective metrics show the superiority of the developed method over several recently developed methods.**

## I. INTRODUCTION

THE Earth observation satellites capture the earth's surface information in different modalities including spectral, spatial and temporal. In the spectral modality, MultiSpectral (MS) data or images are captured at different wavelengths with low spatial resolutions. In the spatial modality, PANchromatic (PAN) data or images are captured over a long range of wavelengths with high spatial resolutions. The fusion of MS and PAN data or multispectral image fusion, named pansharpening in the remote sensing literature [1-8], involves combining the spatial and spectral modalities. In multispectral image fusion, the objective is to recover higher spatial resolutions for multispectral images by reducing the degradation processes that occur during data collection. This recovery can be viewed as an inverse image processing problem. This inverse problem is ill-posed, meaning that there exists a large number of high-resolution images that can get mapped to low-resolution input images. Image fusion techniques attempt to narrow down the search towards obtaining proper high-resolution images.

A number of review articles [9, 10] have categorized pansharpening or multispectral image fusion methods into three main groups consisting of (1) Component Substitution (CS) methods, e.g. [11-13], (2) Multi-Resolution Analysis (MRA) methods, e.g. [14-16], and (3) Model-Based (MB) methods, e.g. [17-19]. The main difference between the first two groups is in the way detail map computation is done. In CS methods, a detail map is acquired by subtracting a PAN image from a linear/nonlinear combination of Low Resolution MS (LRMS) images, whereas in MRA methods, this map is obtained by subtracting a Low Resolution PAN (LRPAN) image from the PAN image. A LRPAN can be computed by applying decomposition methods such as wavelet transform of the PAN image. The third group of pansharpening methods involve using a Bayesian model and posing the fusion as an optimization problem.

In the last few years, deep learning models have been applied to multispectral image fusion generating better outcomes than conventional methods. An initial attempt was made in [20] by solving the pansharpening problem via a deep neural network framework where the nonlinear relationship between the low-resolution and high-resolution images was formulated as a denoising autoencoder. In [21], a three-layer convolutional neural network was designed to turn the MS image fusion problem into a super-resolution problem. The concept of residual learning in MS image fusion was first introduced in [22], where a deep convolutional neural network was used. In [23], a deep denoising auto-encoder approach was utilized to model the relationship between a lowpass version of the PAN image and its high-resolution version. In [24], a deep convolutional neural network was developed for image super-resolution (known as SRCNN), which showed superior performance compared to several other methods. In [25], a pansharpening method was introduced by using the SRCNN as a pre-processing step. In [26], a network structure (known as PanNet) was developed by incorporating the prior knowledge of pansharpening towards enhancing the generalization capability and performance of the network. A Generative Adversarial Networks (GANs) method (known as PSGAN) was discussed in [27] by minimizing the loss between the generative and discriminator parts. One of the advantages of GANs is that it reduces the blurriness on the fused image. Not only it attempts to decrease the $L_1$ loss associated with each pixel, but it also attempts to minimize the loss across the entire fused image.

As far as the loss function in deep neural networks is concerned, a new perceptual loss was presented in [28] to better preserve the spectral information in fused images. In [29], a number of objective functions were examined. In the recently

A. Azarang and N. Kehtarnavaz are with Department of Electrical and Computer Engineering, University of Texas at Dallas. {azarang, kehtar}@utdallas.edu

developed deep learning-based methods, the focus is placed on the preservation of spatial details. For example, the CNN model in [30] was designed for preserving details via a cross-scale learning algorithm. To address the effect of gradient vanishing, the concept of dense connection to pansharpening was extended in [31]. Most of the recently developed deep learning-based methods simply train and regularize the parameters of a network by minimizing a spectral loss between the network output and a pseudo Ground Truth (GT) image. The methods mentioned above primarily use a single objective learning to optimize network parameters and generalize its capability. However, other metrics that can represent both modalities (spatial and spectral) have recently gained more attention. For instance, in [32], based on the correlation maps between MS target images and PAN input images, a loss function was designed to minimize the artifacts of fused images. Also, in [33], it was shown that although a linear combination of MS bands could be estimated from the PAN image, a rather large difference in luminance was resulted. Thus, certain objects could not be differentiated properly. To address this issue, a color-aware perceptual (CAP) loss was designed to obtain the features of a pre-trained VGG network that were more sensitive to spatial details and less sensitive to color differences. The aforementioned methods rely on the availability of GT data for regularizing the network parameters. However, in practice, such data are not available [27].

The objective in this paper is to ease the above two limitations of the existing deep learning models for remote sensing image fusion. The first limitation involves dependency of GT data towards training a network and the second limitation involves the use of a generic loss function. The first limitation is eased by an unsupervised learning strategy based on generative adversarial networks. What is meant by unsupervised learning here is that the label/reference target is not available for training the deep model. A key point for unsupervised learning is that while the data passed through the deep model are abundant, the targets and labels are quite sparse or even non-existent. The second limitation is eased by designing a multi-objective loss function to reflect both spatial and spectral attributes at the same time.

Section II presents the formulation of the developed generative model method as well as its architecture. Section III covers the datasets and objective metrics used in this paper. The experimental results covering both objective metrics and visual comparisons are then provided in Section VI. Finally, the paper is concluded in Section V.

## II. Generative Model Method

This section provides a description of the developed generative model method. To set the stage, let us begin with the general framework of CS methods. The CS framework can be mathematically expressed by the following equation:

$$\hat{\mathbf{M}}_k = \tilde{\mathbf{M}}_k + g_k(\mathbf{P} - \mathbf{I}_k)$$

where $\hat{\mathbf{M}}_k$ and $\tilde{\mathbf{M}}_k$ denote the high-resolution and upsampled low-resolution MS images, respectively, $g_k$'s are injection gains for spectral bands, $\mathbf{P}$ denotes the PAN image, and $\mathbf{I}_k$ is the $k$-th intensity component defined as

$$\mathbf{I}_k = F(\tilde{\mathbf{M}}_k)$$

where $F(.)$ is a linear/nonlinear combination of spectral bands [1-5].

### A) Generative Adversarial Networks

The use of Generative Adversarial Networks (GANs) has been steadily growing. Furthermore, these networks have facilitated the recognition of new categories of learning schemes yielding to the synthesizing of realistic data [34]. In the setting of the GAN structure, rather than a single deep neural network (DNN), training encompasses two DNNs, a "generator" and a "discriminator" architecture, where the former synthesizes realistic data given an input, and the latter classifies inputs as real or synthetic.

In the original form of the GAN architecture [34], the generator is initialized with randomized input noise yielding to several realizations depending on the noise statistics. For image enhancement problems, a specific type of GANs, called conditional GANs (cGANs), is developed since the input to the generator is the image itself, while it could be dissimilar from the output such as an edge map [35]. A noteworthy paper that shows the abilities of GAN in inverse image processing problems is the super-resolution GAN (SRGAN) architecture.

The GAN architecture for the inverse image processing problem here involves an iterative training learning process that alternates between synthesizing of a high-quality image $\mathbf{M}^S$ given a low-quality input image $\mathbf{M}^{IN}$, performed by the generator $G$, and the classification of the high-quality image as real $\mathbf{M}^R$ or synthetic $\mathbf{M}^S$, performed by the discriminator $D$. Thus, training a GAN translates into optimizing a min-max problem where the aim is to estimate the network parameters (weights and biases) of the generator $\theta_G$ and the discriminator $\theta_D$ based on the following equation

$$\min_{\theta_G} \max_{\theta_D} \mathbb{E}[\log D_{\theta_D}(\mathbf{M}^R)] + \mathbb{E}\left[\log\left(1 - D_{\theta_D}\left(G_{\theta_G}(\mathbf{M}^{IN})\right)\right)\right]$$

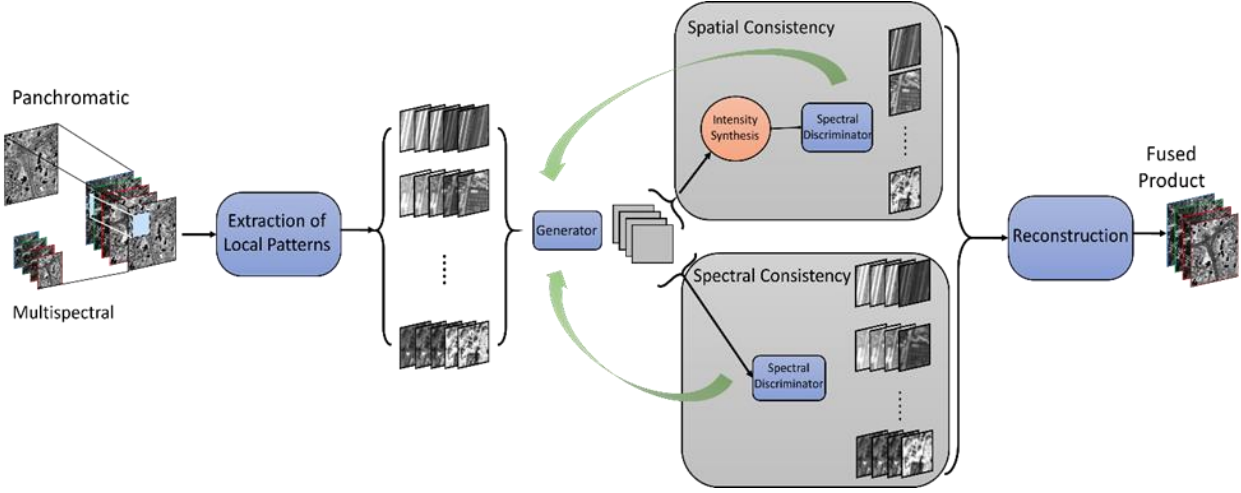

Figure 1. Flowchart of the developed deep generative model for pansharpening.

## B) Fusion Framework

A main contribution of this paper is to formulate the fusion problem as a multi-objective loss function represented by a deep generative network, in which both the spectral and spatial distortions are minimized simultaneously.

***Spectral Preservation Learning Process***: For minimizing the spectral distortion in the fused image, a spectral metric is used to deal with spectral consistency. For this purpose, a discriminator for the learning process is considered, named spectral discriminator here. The MS image data at the original resolution are used as the input of this discriminator. Initially, the output of the generator is inputted to the spectral discriminator. The following objective function is then used to minimize the spectral distortion of the fused image:

$$\mathcal{L}_1 = \mathbb{Q}\big(\widehat{\mathbf{M}}_k^{\mathrm{E}}, \mathbf{M}_k\big)$$

where $\mathbb{Q}(.,.)$ is the Universal Image Quality Index as described in [36], and $\widehat{\mathbf{M}}_k^{\mathrm{E}}$ and $\mathbf{M}_k$ are the estimated high-resolution MS image at the output of the generator and the MS input image at the original resolution, respectively.

***Spatial Preservation Learning Process***: Another discriminator is considered for the minimization of spatial distortion, named spatial discriminator here. To inject spatial details into the fused image and by noting that the PAN image denotes the reference spatial information, the PAN image at the original resolution is used as the input to the discriminator. The following loss function is then used during the training phase of the generative model:

$$\mathcal{L}_2 = \mathbb{Q}\big(\widehat{\mathbf{I}}_k^{\mathrm{E}}, \mathbf{P}_k\big)$$

where $\widehat{\mathbf{I}}_k^{\mathrm{E}}$ is the linear combination of estimated high-resolution MS images at the output of the generator and $\mathbf{P}_k$ is the histogram matched PAN image with respect to the $k$-th spectral band. The learning process of the developed method is illustrated in Fig. 1.

## III. DATASETS AND OBJECTIVE METRICS

For our experimental studies, the following three public domain datasets are used: Pleiades-1A, WorldView-2, and GeoEye-1. The corresponding geographical areas for each dataset are listed in Table I. The MS data for each dataset has four different bands including Blue (B), Green (G), Red (R), and Near InfraRed (NIR). Since the original datasets are quite large, they are divided into 1024×1024 and 256×256 subimages for PAN and MS, respectively. Sample images of each dataset are shown in Figure 2. Note that the regions are selected from different surface indices, e.g. coastal, urban, and jungle areas.

*A) **Reduced Resolution Metrics***: One of the widely used metrics at full-reference mode is Spectral Angle Mapper (SAM) [9]. The color differences between the fused and MS images are characterized by this metric. There are local and global SAM values. Local values are computed as a map via the angle difference between each pixel of the fused image and its corresponding pixel in the MS image. Then, the difference map values are linearized between 0 and 255. The following equation is used to compute local SAM values:

$$\mathrm{SAM}(x, y) = \frac{\langle \mathbf{F}, \mathbf{M} \rangle}{\|\mathbf{F}\|_2 \|\mathbf{M}\|_2}$$

where $\mathbf{F}$ and $\mathbf{M}$ are the pixels of the fused image and the original MS image, respectively. The global value of SAM is computed by taking the average of all the pixels in the SAM map. It is represented in degree (°) or radian. The optimal value for global SAM is zero which means no color distortion in the fused image. Note that SAM is regarded as a spectral distortion metric.

Another objective metric that is widely used is Correlation Coefficient (CC). This metric reflects the cross correlation between the fused and reference images. The range for CC is [-1, 1], where 1 means the highest correlation between images. This metric is computed as follows:

$$CC = \frac{\sum_{x=1}^{N}\sum_{y=1}^{N}(\mathbf{F}(x,y)-\mu_{\mathbf{F}})\,(\mathbf{M}(x,y)-\mu_{\mathbf{M}})}{\sqrt{\sum_{x=1}^{N}\sum_{y=1}^{N}(\mathbf{F}(x,y)-\mu_{\mathbf{F}})^2\,(\mathbf{M}(x,y)-\mu_{\mathbf{M}})^2}}$$

Universal Image Quality Index (UIQI) in [36] is also a widely used metric. This metric denotes a similarity index which characterizes spectral and spatial distortions, and it is computed using the following equation:

$$UIQI = \frac{\sigma_{\mathbf{F,M}}}{\sigma_{\mathbf{F}}\sigma_{\mathbf{M}}}\frac{2\mu_{\mathbf{F}}\mu_{\mathbf{M}}}{\mu_{\mathbf{F}}^2+\mu_{\mathbf{M}}^2}\frac{2\sigma_{\mathbf{F}}\sigma_{\mathbf{M}}}{\sigma_{\mathbf{F}}^2+\sigma_{\mathbf{M}}^2}$$

in which the term $\sigma(.)$ represents the standard deviation and $\mu(.)$ denotes the average. The Q4 metric is a vectorized version of UIQI metric.

The last metric used here in the reduced-resolution mode is Erreur Relative Globale Adimensionnelle de Synthése (ERGAS), which is an improvement of the Mean Squared Error (MSE) by taking into consideration the scale ratio of the PAN and MS images. It reflects the global distortion in the fused image according to the following equation:

$$ERGAS = 100\frac{d_h}{d_l}\sqrt{\frac{1}{N}\sum_{i=1}^{N}\frac{RMSE(\mathbf{F},\mathbf{M})}{\mu(i)}}$$

where $\frac{d_h}{d_l}$ denotes the ratio between pixel sizes of PAN and MS, e.g. $\frac{1}{4}$ for Pleiades-1A, WorldView-2, and GeoEye-1.

**B) Full Resolution Metrics:** Two widely used metrics that quantify the spectral and spatial distortions in the full-resolution mode are $D_\lambda$ and $D_s$. The metric $D_\lambda$ is computed between the low-resolution MS image and the fused image at the PAN image. The UIQI metric between the MS bands, e.g. $\mathbf{M}_1$ and $\mathbf{M}_2$, is first computed and then subtracted from the corresponding multiplication at high-resolution (fused image, i.e. $\mathbf{F}_1$ and $\mathbf{F}_2$). This metric is computed using the following equation:

$$D_\lambda = \sqrt[p]{\frac{1}{N(N-1)}\sum_{i=1}^{N}\sum_{j=1,j\neq i}^{N}\left|UIQI(\mathbf{M}_i,\mathbf{M}_j)-UIQI(\mathbf{F}_i,\mathbf{F}_j)\right|^p}$$

The exponent $p$ is set to one by default but can be chosen to show larger differences between the two terms. Low $D_\lambda$ metric values indicate less spectral distortion and the ideal value is zero.

The metric $D_s$ represents spatial distortion consisting of two terms. The first term is computed at low resolution between the UIQI of the original MS image and the degraded PAN image at the MS resolution and the second term is computed at the PAN resolution between the UIQI of the fused image and the original PAN image. This metric is computed by the following equation:

$$D_s = \sqrt[q]{\frac{1}{N}\sum_{i=1}^{N}\left|UIQI(\mathbf{M}_i,\mathbf{P}^L)-UIQI(\mathbf{F}_i,\mathbf{P})\right|^q}$$

The exponent $q$ is set to one by default. The ideal value for $D_s$ is zero which denotes no spatial distortion.

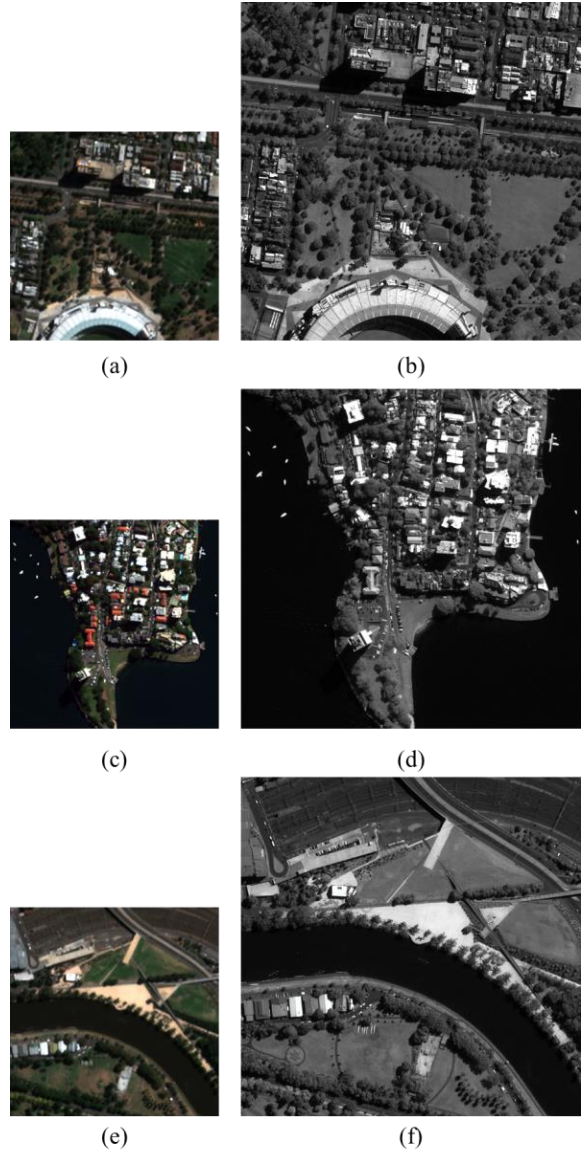

Figure 2. Sample PAN-MS image pairs for three datasets: Pleiades-1A; (a) MS, (b) PAN, WorldView-2; (c) MS, (d) PAN, and GeoEye-1; (e) MS, (f) PAN.

## IV. Experimental Results and Discussion

In this section, the results of the extensive experimentations carried out to examine the developed generative model are reported. Two commonly used protocols of reduced-resolution and high-resolution were considered. The results of the developed method are compared with seven recently developed methods including Band Dependent Spatial Detail (BDSD) [9], Adaptive Intensity-Hue-Saturation (AIHS) [10], CS-based deep learning model (abbreviated as DNN) [20], Convolutional

AutoEncoder-based pansharpening (abbreviated as CAE) [6], Modulation Transfer Function Generalized Laplacian Pyramids (MTF-GLP) [9], Fractional-order Differentiation in Image Fusion (FDIF) [1], our previously developed (Multi-Objective) method [37]. All the experiments were done in both reduced and full-scale modes.

Tables II through VII exhibit the results for the developed method as well as the above representative existing methods. As can be seen from these tables, the developed method provided better metric values. In particular, the SAM and ERGAS metric values were considerably better. These metrics denote the spectral and overall distortions of the fused image, respectively. For the other metrics, the developed method also provided better values in comparison with other methods. Sample fused images for the three datasets examined are shown in Figs. 3 through 6.

Form a visual inspection perspective, one can see that the developed method performed better in terms of both the spatial and spectral contextual information. For example, as can be seen from Fig. 3, the color information of the green area was better preserved. Moreover, as seen from Fig. 5, the spatial details of the PAN image was more effectively injected into the fused image. To make the visual inspection easier, the harbor area in the center of Fig. 5 is magnified. It can be seen that the AIHS method generated color distortion especially in the green area. The MTF-GLP method generated a blurry outcome at the harbour edges. The green area in the FDIF image turned into dark green in comparison to the green color of the LRMS image. The BDSD, DNN, and CAE methods oversharpened the fused image with a slight color distortion. The Multi-Objective method visually produced similar outcomes but the developed method preserved the spectral information better. In order to make the spatial distortion comparison easier, the combination of the NIR-R-G channels of Pleiades-1A is shown in Fig. 4. Visually, in the fused image in this figure, one can see the yellow region on the left side better in terms of spatial details when using the developed generative model. Another example is shown in Fig. 6. In this figure, one can see that the method AIHS suffered from the spectral distortion in some regions. The building edges when using the DNN, CAE, and FDIF methods appeared blurred. The color information in the MTF-GLP and BDSD methods were lost in some regions. The GLP-HRI method suffered from oversharpening. The colors associated with the developed generative model appeared better preserved across different datasets in comparison to our previous (Multi-Objective) method.

TABLE II
Average Performance of Reduced-Resolution Mode of Pleiades-1A Dataset

| | SAM | ERGAS | CC | Q4 |
|---|---|---|---|---|
| BDSD | 3.66 | 4.12 | 0.96 | 0.86 |
| AIHS | 3.24 | 3.86 | 0.96 | 0.87 |
| DNN | 3.15 | 3.82 | 0.96 | 0.92 |
| CAE | 3.05 | 3.75 | 0.96 | 0.92 |
| MTF-GLP | 3.63 | 3.95 | 0.95 | 0.87 |
| FDIF | 3.43 | 4.01 | 0.95 | 0.86 |
| Multi-Objective | 3.03 | 2.79 | 0.97 | 0.93 |
| Developed | **2.53** | **2.36** | **0.97** | **0.94** |
| Ideal | 0 | 0 | 1 | 1 |

TABLE III
Average Performance of Reduced-Resolution Mode of WorldView-2 Dataset

| | SAM | ERGAS | CC | Q4 |
|---|---|---|---|---|
| BDSD | 4.32 | 2.12 | 0.95 | 0.87 |
| AIHS | 4.14 | 1.86 | 0.95 | 0.88 |
| DNN | 3.83 | 1.73 | 0.97 | 0.90 |
| CAE | 3.55 | 1.70 | 0.97 | 0.91 |
| MTF-GLP | 4.60 | 2.34 | 0.94 | 0.86 |
| FDIF | 4.43 | 2.30 | 0.93 | 0.85 |
| Multi-Objective | 3.30 | 1.64 | 0.97 | 0.92 |
| Developed | **3.16** | **1.41** | **0.97** | **0.92** |
| Ideal | 0 | 0 | 1 | 1 |

TABLE IV
Average Performance of Reduced-Resolution Mode of GeoEye-1 Dataset

| | SAM | ERGAS | CC | Q4 |
|---|---|---|---|---|
| BDSD | 2.76 | 1.88 | 0.93 | 0.86 |
| AIHS | 2.34 | 1.78 | 0.93 | 0.88 |
| DNN | 2.25 | 1.64 | 0.94 | 0.90 |
| CAE | 2.15 | 1.55 | 0.94 | 0.90 |
| MTF-GLP | 2.60 | 2.02 | 0.93 | 0.86 |
| FDIF | 2.43 | 2.12 | 0.93 | 0.88 |
| Multi-Objective | 2.03 | 1.55 | 0.97 | 0.93 |
| Developed | **1.84** | **1.30** | **0.97** | **0.95** |
| Ideal | 0 | 0 | 1 | 1 |

TABLE V
Average Performance of Full-Resolution Mode of
Pleiades-1A Dataset

|  | $D_s$ | $D_\lambda$ | QNR |
|---|---|---|---|
| BDSD | 0.10 | 0.10 | 0.81 |
| AIHS | 0.11 | 0.12 | 0.78 |
| DNN | 0.08 | 0.07 | 0.86 |
| CAE | 0.06 | 0.06 | 0.88 |
| MTF-GLP | 0.10 | 0.12 | 0.78 |
| FDIF | 0.09 | 0.10 | 0.82 |
| Multi-objective | 0.06 | 0.05 | 0.89 |
| Developed | **0.04** | **0.04** | **0.92** |
| Ideal | 0 | 0 | 1 |

TABLE VI
Average Performance of Full-Resolution Mode of
WorldView-2 Dataset

|  | $D_s$ | $D_\lambda$ | QNR |
|---|---|---|---|
| BDSD | 0.11 | 0.06 | 0.84 |
| AIHS | 0.12 | 0.06 | 0.83 |
| DNN | 0.07 | 0.05 | 0.88 |
| CAE | 0.06 | 0.05 | 0.89 |
| MTF-GLP | 0.09 | 0.12 | 0.80 |
| FDIF | 0.10 | 0.09 | 0.82 |
| Multi-objective | 0.05 | 0.05 | 0.90 |
| Developed | **0.03** | **0.04** | **0.93** |
| Ideal | 0 | 0 | 1 |

TABLE VII
Average Performance of Full-Resolution Mode of
GeoEye-1 Dataset

|  | $D_s$ | $D_\lambda$ | QNR |
|---|---|---|---|
| BDSD | 0.08 | 0.03 | 0.89 |
| AIHS | 0.12 | 0.08 | 0.81 |
| DNN | 0.05 | 0.02 | 0.93 |
| CAE | 0.04 | 0.02 | 0.94 |
| MTF-GLP | 0.09 | 0.14 | 0.78 |
| FDIF | 0.12 | 0.10 | 0.79 |
| Multi-objective | 0.03 | 0.02 | 0.95 |
| Developed | **0.02** | **0.01** | **0.97** |
| Ideal | 0 | 0 | 1 |

## V. Conclusion

In this paper, a new generative model method for unsupervised learning process of multispectral image fusion has been developed. The developed method addresses the ill-posed pansharpening problem in a more comprehensive manner. The model consists of two separate discriminators for learning the spectral and spatial information. The former uses the MS data at the original scale as the input of the spectral discriminator. The latter uses the PAN image as the input. A comprehensive comparison has been conducted with seven recent pansharpening methods and the results obtained show fused images are generated with less distortion compared to these methods.

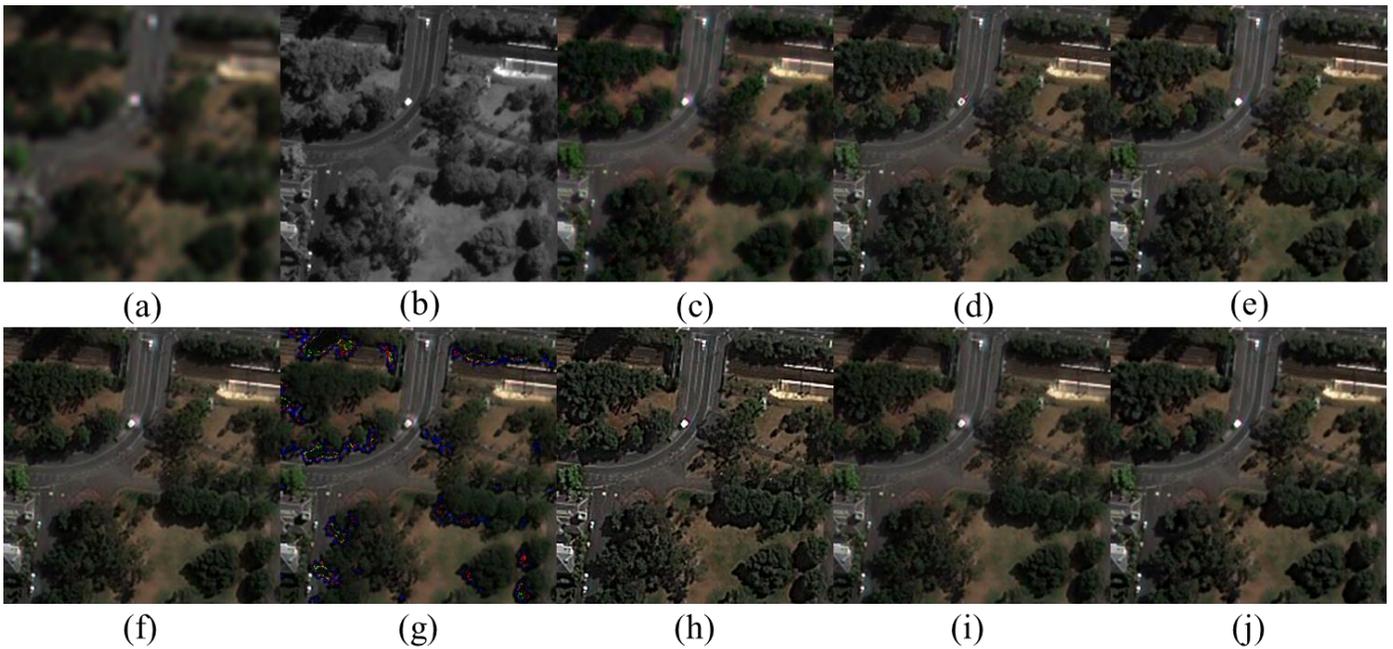

Figure 3. True color representation of sample fusion results for Pleiades-1A dataset: (a) MS, (b) PAN, (c) BDSD, (d) AIHS, (e) DNN, (f) CAE, (g) MTF-GLP, (h) FDIF, (i) Multi-Objective, (j) Developed.

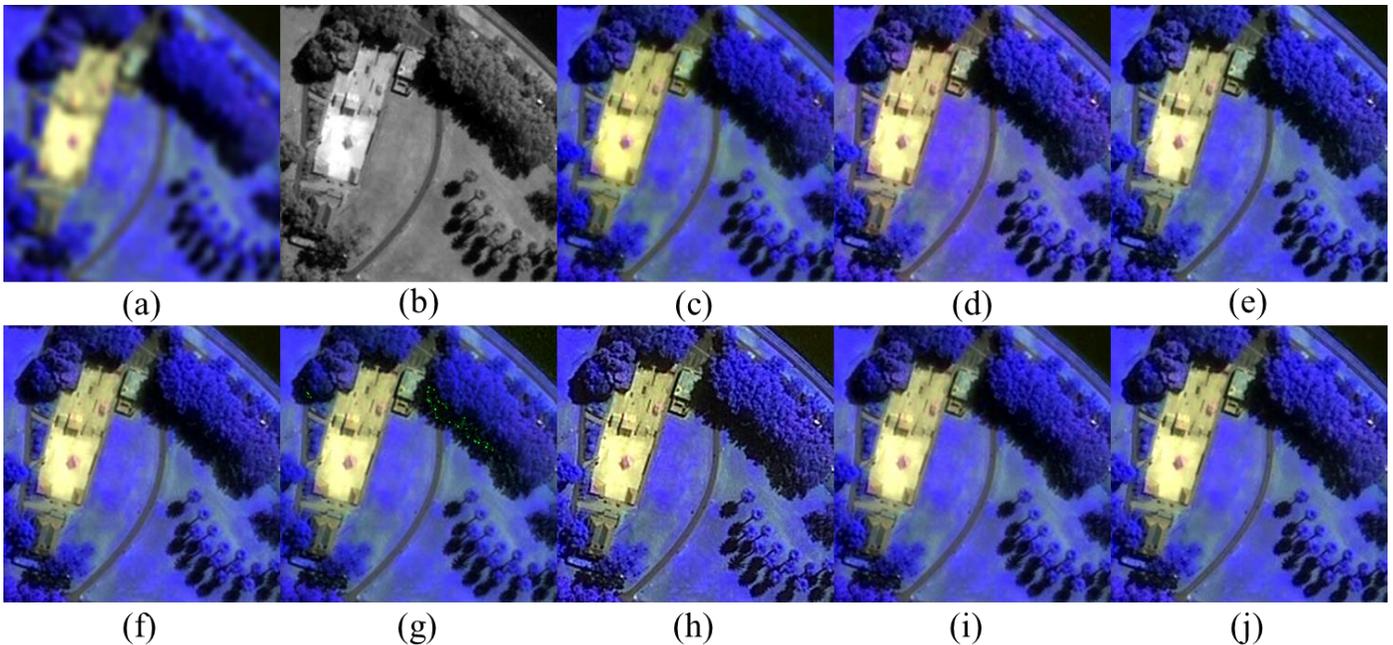

Figure 4. False color representation of sample fusion results for Pleiades-1A dataset: (a) MS, (b) PAN, (c) BDSD, (d) AIHS, (e) DNN, (f) CAE, (g) MTF-GLP, (h) FDIF, (i) Multi-Objective, (j) Developed.

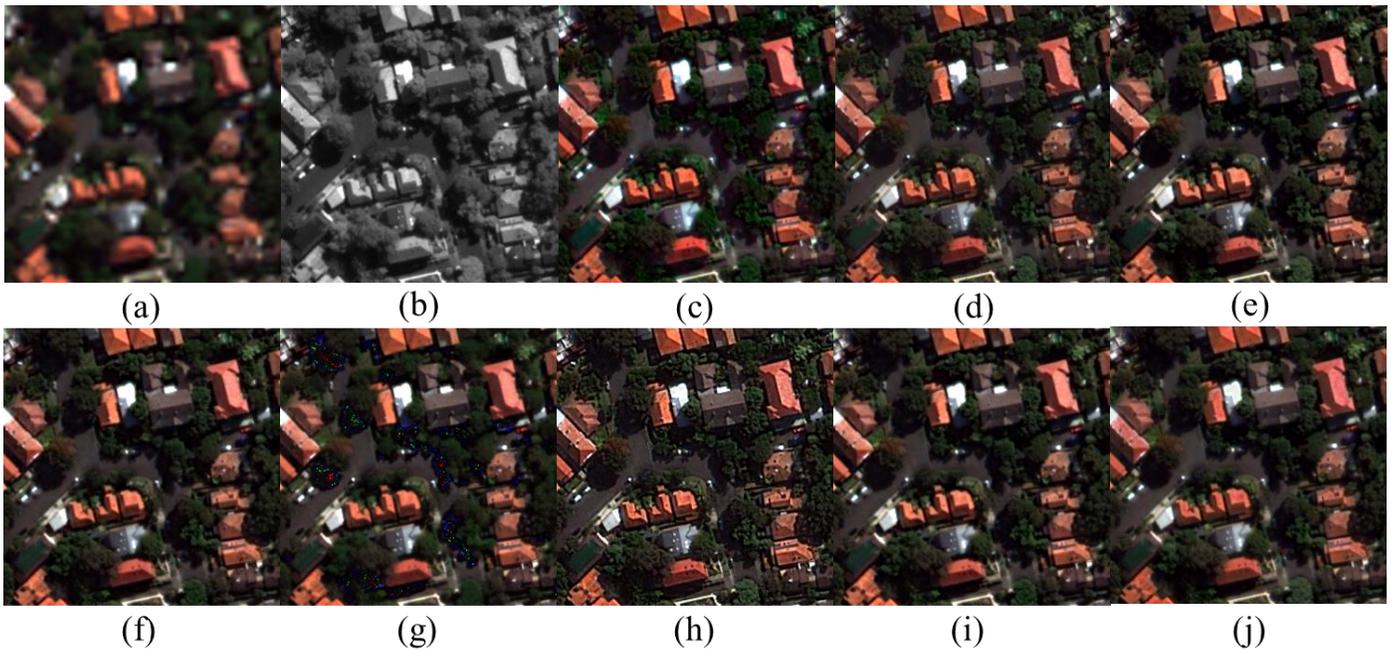

Figure 5. True color representation of sample fusion results for WorldView-2 dataset: (a) MS, (b) PAN, (c) BDSD, (d) AIHS, (e) DNN, (f) CAE, (g) MTF-GLP, (h) FDIF, (i) Multi-Objective, (j) Developed.

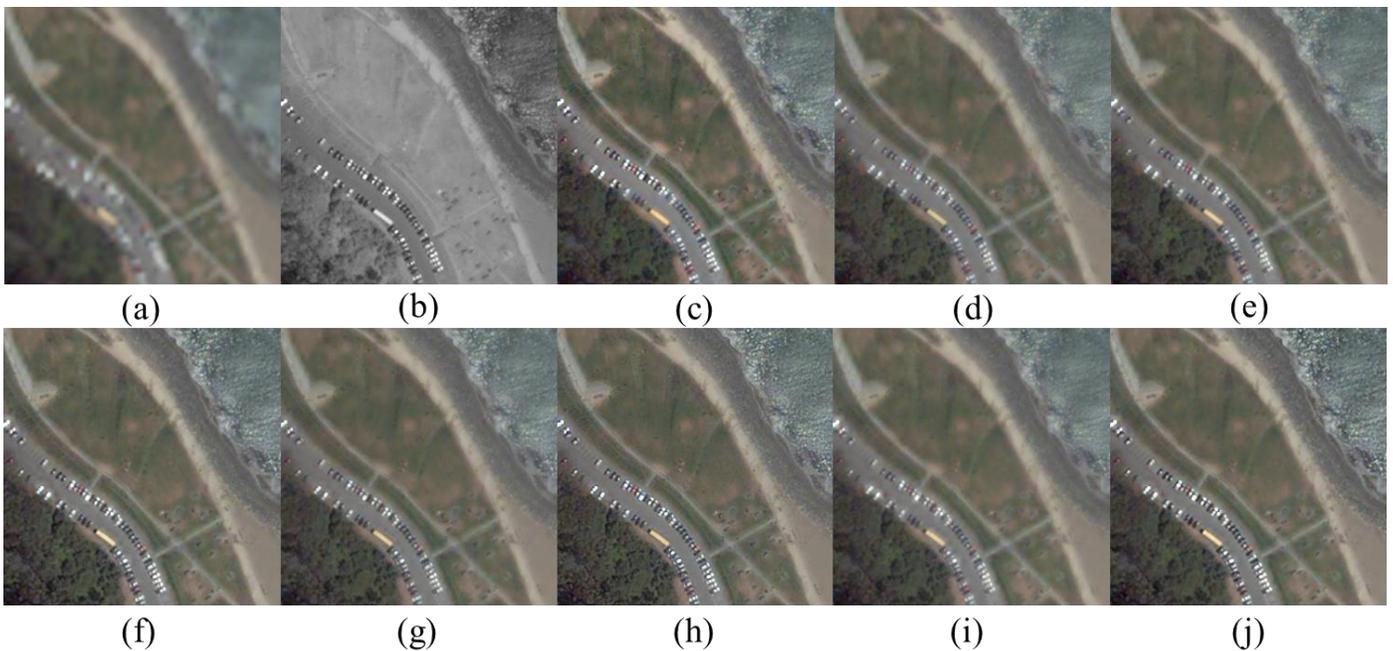

Figure 6. True color representation of sample fusion results for GeoEye-1 dataset: (a) MS, (b) PAN, (c) BDSD, (d) AIHS, (e) DNN, (f) CAE, (g) MTF-GLP, (h) FDIF, (i) Multi-Objective, (j) Developed.